\documentclass[journal]{IEEEtran}
\usepackage{latexsym,amssymb,amsmath,graphicx,epsfig,cite,bbm,float,epsf,graphics}

\newcommand{\sy}{\left\{y(t)\right\}_{t \geq 1}}

\newcommand{\be}{\begin{equation}}
\newcommand{\ee}{\end{equation}}
\newcommand{\bea}{\begin{eqnarray}}
\newcommand{\eea}{\end{eqnarray}}
\newcommand{\nn}{\nonumber}

\newcommand{\MB}{\left[\begin{array}}
\newcommand{\ME}{\end{array}\right]}

\renewcommand{\vec}[1]{\mbox{\boldmath${#1}$}}

\newcommand{\vu}{\vec{u}}
\newcommand{\vw}{\vec{w}}
\newcommand{\ei}{\end{itemize}}
\newcommand{\bi}{\begin{itemize}}

\newcommand{\vx}{\mbox{$\vec{x}$}}

\newcommand{\lp}{\lambda^+}

\newcommand{\defi}{\stackrel{\bigtriangleup}{=}}

\begin{document}
\title{A New Analysis of an Adaptive Convex Mixture: A Deterministic Approach}
\author{Mehmet~A.~Donmez, Sait Tunc and Suleyman~S.~Kozat,~\IEEEmembership{Senior Member}
\thanks{ This work is supported in part by IBM Faculty Award and Outstanding Young Scientist Award Program, Turkish Academy of Sciences. Suleyman S. Kozat, Mehmet A. Donmez and Sait Tunc (\{skozat,medonmez,saittunc\}@ku.edu.tr) are with the Competitive Signal Processing Laboratory at Koc University, Istanbul, tel: +902123381864.}}

\maketitle

\begin{abstract}
We introduce a new analysis of an adaptive mixture method that
combines outputs of two constituent filters running in parallel to
model an unknown desired signal.  This adaptive mixture is shown to
achieve the mean square error (MSE) performance of the best
constituent filter, and in some cases outperforms both, in the
steady-state. However, the MSE analysis of this mixture in the
steady-state and during the transient regions uses approximations and
relies on statistical models on the underlying signals and
systems. Hence, such an analysis may not be useful or valid for
signals generated by various real life systems that show high degrees
of nonstationarity, limit cycles and, in many cases, that are even
chaotic. To this end, we perform the transient and the steady-state
analysis of this adaptive mixture in a ``strong'' deterministic sense
without any approximations in the derivations or statistical
assumptions on the underlying signals such that our results are
guaranteed to hold. In particular, we relate the time-accumulated
squared estimation error of this adaptive mixture at any time to the
time-accumulated squared estimation error of the optimal convex
mixture of the constituent filters directly tuned to the
underlying signal in an individual sequence manner.
\end{abstract}

\IEEEpeerreviewmaketitle

\section{Introduction}
\label{sec:introduction}
The problem of estimating an unknown desired signal is heavily
investigated in the adaptive signal processing literature. However, in
various applications, certain difficulties arise in the estimation
process due to the lack of structural and statistical information
about the data model that relates the observation process to the
desired signal. To resolve this lack of information, mixture
approaches are proposed that adaptively combine outputs of multiple
constituent algorithms performing the same task \cite{sinfed,
  convex,tranconv}. These parallel running algorithms can be seen as
alternative hypotheses for modeling, which can be exploited for both
performance improvement and robustness. Along these lines, a convexly
constrained mixture method that combines outputs of two adaptive
filters is introduced in \cite{convex}. In this approach, the outputs
of the constituent algorithms are adaptively combined under a convex
constraint to minimize the final MSE. This adaptive mixture is shown
to be universal with respect to the input filters in a certain
stochastic sense such that it achieves (and in some cases outperforms)
the MSE performance of the best constituent filter in the mixture in
the steady-state. However, the MSE analysis of this adaptive mixture
for the steady-state and during the transient regions uses
approximations, e.g., separation assumptions, and relies on
statistical models on the signals and systems, e.g., nonstationarity
data models \cite{convex,tranconv,kozat}.

Nevertheless, signals produced by various real life systems, such as
in underwater acoustic communication applications, show high degrees
of nonstationarity, limit cycles and, in many cases, are even chaotic
so that they hardly fit to assumed statistical models. Hence an analysis
based on certain statistical assumptions or approximations may not
useful or adequate under these conditions. To this end, we refrain
from making any statistical assumptions on the underlying signals and
present an analysis that is guaranteed to hold for any bounded
arbitrary signal without any approximations. In particular, we relate
the performance of this adaptive mixture to the performance of the
optimal convex combination that is directly tuned to the underlying
signal and outputs of the constituent filters in a deterministic
sense. Naturally, this optimal convex combination can only be
chosen in hindsight after observing the whole signal and outputs a
priori (before we even start processing the data).  In this sense, we
provide both the transient and steady-state analysis of the adaptive
mixture in a deterministic sense without any assumptions on the
underlying signals or any approximations in the derivations.  Our
results are guaranteed to hold in an individual sequence manner.

After we provide a brief system description in
Section~\ref{sec:problem_description}, we present a deterministic
analysis of the convexly constrained adaptive mixture method in
Section~\ref{sec:deterministic_analysis}, where the performance bounds
are given as a theorem and a lemma. The letter concludes with certain
remarks.

\section{Problem Description}

\label{sec:problem_description}
In this framework, we have a desired signal $\sy$, where $|y(t)| \leq
Y <\infty$, and two constituent filters running in parallel producing
$\{\hat{y}_1(t)\}_{t \geq 1}$ and $\{\hat{y}_2(t)\}_{t \geq 1}$,
respectively, as the estimations (or predictions) of the desired
signal $\sy$. We assume that $Y$ is known. Here, we have no
restrictions on $\hat{y}_1(t)$ or $\hat{y}_2(t)$, e.g., these outputs
are not required to be causal, however, without loss of generality, we
assume $|\hat{y}_1(t)| \leq Y$ and $|\hat{y}_2(t)| \leq Y$, i.e.,
these outputs can be clipped to the range $[-Y,Y]$ without sacrificing
performance under the squared error. As an example, the desired signal
and outputs of the first stage filters can be single realizations
generated under the framework of \cite{convex}. At each time $t$, the
convexly constrained algorithm receives an input vector $\vx(t) \defi
[\hat{y}_1(t)\;\hat{y}_2(t)]^T$ and outputs
\begin{align*}
\hat{y}(t) = \lambda(t)  \hat{y}_1(t) + (1-\lambda(t)) \hat{y}_2(t) = [\lambda(t) \; (1-\lambda(t))] \vx(t),
\end{align*}\normalsize
where $0 \leq \lambda(t) \leq 1$, as the final estimate. The final estimation error is given by $e(t)=y(t) -
\hat{y}(t)$. The combination weight $\lambda(t)$ is trained through an
auxiliary variable using a stochastic gradient update to minimize the
squared final estimation error as
\begin{align}
& \lambda(t)  = \frac{1}{1+e^{-\rho(t)}} \label{eq:son2}, \\
& \rho(t+1)  = \rho(t)-\mu \nabla_{\rho}e^2(t)\big|_{\rho=\rho(t)} \nonumber \\
  & =  \rho(t)+ \mu e(t)\lambda(t)(1-\lambda(t)) [\hat{y}_1(t)-\hat{y}_2(t)], \label{eq:1}
\end{align}\normalsize
where $\mu > 0$ is the learning rate. The combination
parameter $\lambda(t)$ in \eqref{eq:son2} is constrained to lie in
$[\lambda^+,(1-\lambda^+)]$, $0<\lambda^+ < 1/2$ in \cite{convex},
since the update in \eqref{eq:1} may slow down when $\lambda(t)$ is
too close to the boundaries. We follow the same restriction and
analyze \eqref{eq:1} under this constraint.

When applied to any sequence $\sy$, the algorithm of \eqref{eq:son2}
yields the total accumulated loss
\[
L_n(\hat{y},y) \defi \sum_{t=1}^n
(y(t)-\hat{y}(t))^2
\]\normalsize
for any $n$. Although, we use the time-accumulated squared error as
the performance measure, our results can be readily extended to
the exponentially weighted accumulated squared error. We next provide deterministic bounds on $L_n(\hat{y},y)$ with respect to the best convex combination $
\min\limits_{\beta\in[0,1]} L_n(\hat{y}_{\beta},y)$, where
\[
 L_n(\hat{y}_{\beta},y) = \sum_{t=1}^n (y(t)-\hat{y}_{\beta}(t))^2
\]
and $\hat{y}_{\beta}(t)\defi\beta \hat{y}_1(t)+(1-\beta)\hat{y}_2(t)$,
that holds uniformly in an individual sequence manner without any
stochastic assumptions on $y(t)$, $\hat{y}_1(t)$, $\hat{y}_2(t)$ or
$n$. Note that the best convex combination $
\min\limits_{\beta\in[0,1]} L_n(\hat{y}_{\beta},y)$, which we compare
the performance against,
can only be determined after observing the entire sequences, i.e.,
$\{y(t)\},\{\hat{y}_1(t)\}$ and $\{\hat{y}_2(t)\}$, in advance for all $n$.

\section{A Deterministic Analysis \label{sec:deterministic_analysis}}

In this section, we first relate the accumulated loss of the adaptive
mixture to the accumulated loss of the best convex combination that
minimizes the accumulated loss in the following theorem. Then, we
demonstrate that one cannot improve the convergence rate of this upper
bound using our methodology directly and the Kullback-Leibler (KL)
divergence \cite{KiWa02} as the distance measure by providing counter
examples as a lemma. We emphasize that although the steady-state and
transient MSE performances of the convexly constrained mixture
algorithm are analyzed with respect to the constituent filters
\cite{convex,tranconv,kozat}, we perform the steady-state and
transient analysis without any stochastic assumptions or use any
approximations in the following theorem.\\

\noindent
{\bf Theorem:} The algorithm given in \eqref{eq:1}, when applied to
any sequence $\sy$, with $|y(t)| \leq Y<\infty$, yields, for any $n$
and any $\epsilon > 0$
\begin{equation}
\frac{L_n(\hat{y},y)}{n}-\left( \frac{2 \epsilon+1}{1-z^2}\right) \min\limits_{\beta\in[0,1]} \left\{ \frac{L_n(\hat{y}_{\beta},y)}{n}\right\} \leq O\left( \frac{1}{n\epsilon} \right), \label{eq:theorem}
\end{equation}\normalsize
where $\hat{y}_{\beta}(t)=\beta \hat{y}_1(t)+(1-\beta)\hat{y}_2(t)$, $z\defi \frac{1-4 \lambda^+(1-\lambda^+)}{1+4
  \lambda^+(1-\lambda^+)} < 1$ and step size $\mu = \frac{4
  \epsilon}{2\epsilon+1}\frac{2+2z}{Y^2}$, provided that $\lambda(t)
\in \left[\lambda^+ , 1-\lambda^+\right]$, $0 < \lambda^+ < 1/2$, for all
$t$ during the adaptation. \\

Equation \eqref{eq:theorem} provides the exact trade-off between the
transient and steady-state performances of the adaptive mixture in a
deterministic sense without any assumptions or approximations.  From
 \eqref{eq:theorem} we observe that the convergence rate of the right hand
side is $O\left( \frac{1}{n\epsilon} \right)$ and, as in the
stochastic case \cite{kozat}, to get a tighter asymptotic bound with
respect to the optimal convex combination of the filters, we require a
smaller $\epsilon$, i.e., smaller learning rate $\mu$, which increases
the right hand side of \eqref{eq:theorem}.  Although this result is
well-known in the adaptive filtering literature and appears widely in
stochastic contexts, however, this trade-off is guaranteed to hold in
here without any statistical assumptions or approximations.  Note that
the optimal convex combination in \eqref{eq:theorem}, i.e., minimizing $\beta$,
depends on the entire signal and outputs of the
constituent filters for all $n$.\\

\noindent
{\bf Proof:} To prove the theorem, we use the approach introduced in
\cite{cesab} (and later used in \cite{KiWa02}) based on
measuring progress of an adaptive algorithm using certain distance
measures.

We first convert \eqref{eq:1} to a direct update on $\lambda(t)$ and
use this direct update in the proof. Using $e^{-\rho(t)} = \frac{1-\lambda(t)}{\lambda(t)}$ from \eqref{eq:son2}, the update in \eqref{eq:1} can be written as
\small
\begin{align}
&\lambda(t+1)\nn\\ &= \frac{1}{1+e^{-\rho(t+1)}}
              = \frac{1}{1+ \frac{1-\lambda(t)}{\lambda(t)} e^{-\mu e(t)\lambda(t)(1-\lambda(t)) [\hat{y}_1(t)-\hat{y}_2(t)]}} \nonumber \\
             & = \frac{\lambda(t)e^{\mu e(t) \lambda(t) (1-\lambda(t)) \hat{y}_1(t)}}{\lambda(t)e^{\mu e(t) \lambda(t) (1-\lambda(t)) \hat{y}_1(t)} + (1-\lambda(t))e^{\mu e(t) \lambda(t) (1-\lambda(t)) \hat{y}_2(t)}}.
\label{update}
\end{align}\normalsize
Unlike \cite{KiWa02} (Lemma 5.8), our update in \eqref{update} has,
in a certain sense, an adaptive learning rate $\mu \lambda(t)
(1-\lambda(t))$ which requires different formulation, however, follows
similar lines of \cite{KiWa02} in certain parts.

Here, we first define $\hat{y}_{\beta}(t)\defi\beta \hat{y}_1(t)+(1-\beta)\hat{y}_2(t)=\vu^T \vx(t)$, where $\beta\in[0,1]$ and $\vu \defi [\beta\;\;1-\beta]^T$.  At each
adaptation, the progress made by the algorithm towards $\vu$ at time
$t$ is measured as $d(\vu,\vw(t)) - d(\vu,\vw(t+1))$, where $\vw(t)
\defi [\lambda(t) \; (1-\lambda(t))]^T$ and $d(\vu,\vw) \defi
\sum_{i=1}^2 u_i \ln (u_i/w_i)$ is the
Kullback-Leibler divergence \cite{cesab}, $\vu \in \left[0,1\right]^2$, $\vw \in \left[0,1\right]^2$. We require that this
progress is at least $a(y(t)-\hat{y}(t))^2 - b(y(t)-\hat{y}_{\beta}(t))^2$
for certain $a$, $b$, $\mu$ \cite{cesab,KiWa02}, i.e.,
\begin{align}
&a(y(t)-\hat{y}(t))^2 - b(y(t)-\hat{y}_{\beta}(t))^2 \nn\\ &\leq [d(\vu,\vw(t)) - d(\vu,\vw(t+1))] \nn\\
&= \beta \ln\left(\frac{\lambda(t+1)}{\lambda(t)}\right)+(1-\beta) \ln\left(\frac{1-\lambda(t+1)}{1-\lambda(t)}\right),
\label{desired_bound}
\end{align}
which yields the desired deterministic bound in \eqref{eq:theorem}
after telescoping.

Defining $\zeta(t) = e^{\mu e(t)\lambda(t)(1 - \lambda(t))}$, we have from \eqref{update}
\begin{align}
&\beta \ln\left(\frac{\lambda(t+1)}{\lambda(t)}\right)+(1-\beta) \ln\left(\frac{1-\lambda(t+1)}{1-\lambda(t)}\right) \nn\\&= \hat{y}_{\beta}(t) \ln \zeta(t) - \ln(\lambda(t) \zeta(t)^{\hat{y}_1(t)} + (1-\lambda(t))\zeta(t)^{\hat{y}_2(t)}).\label{eq:b}
\end{align}\normalsize
Using the inequality $\alpha^x \leq 1 - x(1-\alpha)$ for $\alpha \geq 0$ and $x \in [0,1]$ from \cite{cesab}, we have
\begin{align*}
\zeta(t)^{\hat{y}_1(t)} & = (\zeta(t)^{2Y})^{\frac{\hat{y}_1(t) + Y}{2Y}} \zeta(t)^{-Y} \\
                   & \leq \zeta(t)^{-Y}\left(1 - \frac{\hat{y}_1(t) + Y}{2Y}(1- \zeta(t)^{2Y})\right), \nonumber
\end{align*}\normalsize
which implies in \eqref{eq:b}
\begin{align}
&\ln\left(\lambda \zeta(t)^{\hat{y}_1(t)} + (1-\lambda) \zeta(t)^{\hat{y}_2(t)}\right) \nn\\
&\leq \ln \left( \zeta(t)^{-Y}(1 - \frac{\lambda \hat{y}_1(t) + (1-\lambda) \hat{y}_2(t) + Y}{2Y}(1- \zeta(t)^{2Y})) \right) \nonumber \\
& = -Y \ln \zeta(t) + \ln \left(1 - \frac{\hat{y}(t)+ Y}{2Y}(1-\zeta(t)^{2Y})\right),\label{eq:ln}
\end{align}\normalsize
where $\hat{y}(t)= \lambda(t) \hat{y}_1(t) + (1-\lambda(t))
\hat{y}_2(t)$. As in \cite{KiWa02}, one can further bound
\eqref{eq:ln} using $\ln(1-q(1-e^p)) \leq pq+\frac{p^2}{8}$ for $0
\leq q<1$ (originally from \cite{cesab})
\begin{align}
&\ln\left(\lambda \zeta(t)^{\hat{y}_1(t)} + (1-\lambda) \zeta(t)^{\hat{y}_2(t)}\right) \nn\\
&\leq  -Y \ln \zeta(t) +  (\hat{y}(t)+ Y) \ln\zeta(t) +\frac{Y^2 (\ln \zeta(t))^2}{2}. \label{eq:a}
\end{align}\normalsize
Using
\eqref{eq:a} in \eqref{eq:b} yields
\begin{align}
&\beta \ln\left(\frac{\lambda(t+1)}{\lambda(t)}\right)+(1-\beta) \ln\left(\frac{1-\lambda(t+1)}{1-\lambda(t)}\right) \geq \label{eq:c}\\ & (\hat{y}_{\beta}(t) + Y)\ln \zeta(t) -  (\hat{y}(t)+ Y) \ln\zeta(t) - \frac{Y^2 (\ln \zeta(t))^2}{2} . \nn
\end{align}\normalsize
From now on, we omit $\beta$ of $\hat{y}_{\beta}(t)$. We observe from \eqref{desired_bound} and \eqref{eq:c} that to prove the
theorem, it is sufficient to show that
$G(y(t),\hat{y}(t),\hat{y}_{\beta}(t),\zeta(t))\leq 0$, where
\begin{align}
&G(y(t),\hat{y}(t),\hat{y}_{\beta}(t),\zeta(t)) \nn\\
& \defi -(\hat{y}_{\beta}(t) + Y)\ln \zeta(t) +  (\hat{y}(t)+ Y) \ln\zeta(t)  \nn \\ & +\frac{Y^2 (\ln \zeta(t))^2}{2}
                                        + a(y(t)-\hat{y}(t))^2 - b(y(t)-\hat{y}_{\beta}(t))^2.\label{main_func}
\end{align}\normalsize
For fixed $y(t),\hat{y}(t),\zeta(t)$,
$G(y(t),\hat{y}(t),\hat{y}_{\beta}(t),\zeta(t))$ is maximized when
$\frac{\partial G}{\partial {\hat{y}_{\beta}(t)}}=0$, i.e.,
$ \hat{y}_{\beta}(t) -
y(t)+ \frac{\ln \zeta(t)}{2b} = 0$
since $\frac{\partial^2 G}{\partial {\hat{y}_{\beta}(t)}^2} = -2b < 0$,
yielding $
{\hat{y}_{\beta}(t)}^* = y(t)- \frac{\ln \zeta(t)}{2b}$.
Note that while taking the partial derivative of $G(\cdot)$ with
respect to $\hat{y}_{\beta}(t)$ and finding ${\hat{y}_{\beta}(t)}^*$, we assume
that all $y(t),\hat{y}(t),\zeta(t)$ are fixed, i.e., their partial
derivatives with respect to $\hat{y}_{\beta}(t)$ is zero. This yields an
upper bound on $G(\cdot)$ in terms of $\hat{y}_{\beta}(t)$. Hence, it is
sufficient to show that $G(y(t),\hat{y}(t),{\hat{y}_{\beta}(t)}^*,\zeta(t))
\leq 0$ such that \cite{KiWa02}
\begin{align}
& G(y(t),\hat{y},{\hat{y}_{\beta}(t)}^*,\zeta(t)) \nonumber \\
& = - \left(y(t)+ Y - \frac{\ln \zeta(t)}{2b}\right) \ln \zeta(t) +   (\hat{y}(t)+ Y) \ln\zeta(t) \nonumber \\
& +\frac{Y^2 (\ln \zeta(t))^2}{2}  + a(y(t)-\hat{y}(t))^2 - \frac{(\ln \zeta(t) )^2}{4b}\label{eq:ust1} \\
& = a (y(t)-\hat{y}(t))^2 - (y(t)-\hat{y}(t))\ln \zeta(t) + \frac{(\ln \zeta(t))^2}{4b} \nn \\
&+ \frac{Y^2 (\ln \zeta(t))^2}{2}\nonumber\\
& = (y(t) - \hat{y}(t))^2\times \Bigg[  a - \mu\lambda(t)(1-\lambda(t))   \nn \\
&+\frac{{\mu}^2{\lambda(t)}^2 (1-\lambda(t))^2}{4b} + \frac{Y^2 {\mu}^2 {\lambda(t)}^2 (1-\lambda(t))^2}{2} \Bigg]. \label{eq:last}
\end{align}\normalsize

For \eqref{eq:last} to be negative, defining $k \defi \lambda(t) (1-
\lambda(t))$ and
\[
H(k) \defi k^2 \mu^2 (\frac{Y^2}{2} + \frac{1}{4b}) -
\mu k + a,
\]\normalsize
it is sufficient to show that $H(k) \leq 0$ for $k \in [\lp (1-\lp) ,
  \frac{1}{4}]$, i.e., $ k \in [\lp (1-\lp) , \frac{1}{4}]$ when
$\lambda(t) \in [\lp,(1-\lp)]$, since $H(k)$ is a convex quadratic
function of $k$, i.e., $\frac{\partial^2 H}{\partial k^2} > 0$. Hence,
we require the interval where the function $H(\cdot)$ is negative
should include $[\lp(1-\lp),\frac{1}{4}]$, i.e., the roots $k_1$ and
$k_2$ (where $k_2 \leq k_1$) of $H(\cdot)$ should satisfy $k_1 \geq
\frac{1}{4}$ and $k_2 \leq \lp(1-\lp)$, where
\begin{align}
k_{1,2} & = \frac{\mu \pm \sqrt{{\mu}^2 - 4 {\mu}^2 a \left(\frac{Y^2}{2} + \frac{1}{4b}\right)}}{2{\mu}^2 (\frac{Y^2}{2} + \frac{1}{4b})} \nn \\
& = \frac{1 \pm \sqrt{1 - 4  a s}}{2\mu s}\label{eq:root1}
\end{align}and \[s \defi \left(\frac{Y^2}{2} + \frac{1}{4b}\right)\].

To satisfy $k_1 \geq 1/4$, we straightforwardly require from \eqref{eq:root1}
\[
\frac{2+2 \sqrt{1-4as}}{s} \geq
\mu.
\]\normalsize
To get the tightest upper bound for \eqref{eq:root1}, we set
\[
\mu = \frac{2+2 \sqrt{1-4as}}{s},
\]\normalsize
i.e., the largest allowable learning rate.

To have $k_2 \leq
\lambda^+(1-\lambda^+)$ with $\mu = \frac{2+2 \sqrt{1-4as}}{s}$, from
\eqref{eq:root1} we require \be
\frac{1-\sqrt{1-4as}}{4(1+\sqrt{1-4as})} \leq
\lambda^+(1-\lambda^+). \label{eq:yeter} \ee Equation \eqref{eq:yeter}
yields
\begin{equation}
as= a \left(\frac{Y^2}{2} + \frac{1}{4b}\right) \leq \frac{1- z^2}{4}, \label{eq:3}
\end{equation}\normalsize
where
\[
z \defi \frac{1-4 \lambda^+(1-\lambda^+)}{1+4
  \lambda^+(1-\lambda^+)}
\]\normalsize
and $z < 1$ after some algebra.

To satisfy \eqref{eq:3}, we set $b = \frac{\epsilon}{Y^2}$ for any (or
arbitrarily small) $\epsilon > 0$ that results
\begin{equation}
a \leq \frac{(1-z^2) \epsilon}{Y^2 (2\epsilon+1)}. \label{eq:4}
\end{equation}\normalsize
To get the tightest bound in \eqref{desired_bound}, we select $a =
\frac{(1-z^2) \epsilon}{Y^2 (2\epsilon+1)}$ in \eqref{eq:4}. Such
selection of $a$, $b$ and $\mu$ results in \eqref{desired_bound}

\begin{align}
&\left(\frac{(1-z^2) \epsilon}{Y^2 (2\epsilon+1)}\right)
(y(t)-\hat{y}(t))^2 - \left( \frac{\epsilon}{Y^2}\right)
(y(t)- \hat{y}_{\beta}(t))^2  \nn\\&\leq
\beta \ln\left(\frac{\lambda(t+1)}{\lambda(t)}\right)+(1-\beta) \ln\left(\frac{1-\lambda(t+1)}{1-\lambda(t)}\right). \label{eq:fin1}
\end{align}\normalsize
After telescoping, i.e., summation over $t$, $\sum_{t=1}^n$,
\eqref{eq:fin1} yields
\small
\begin{align}
&aL_n(\hat{y},y)-b \min\limits_{\beta\in[0,1]} \left\{ \frac{L_n(\hat{y}_{\beta},y)}{n}\right\} \nn\\
&\leq \beta \ln\left(\frac{\lambda(t+1)}{\lambda(1)}\right)+(1-\beta) \ln\left(\frac{1-\lambda(t+1)}{1-\lambda(1)}\right)  \leq  O(1), \nonumber \\
&\left(\frac{(1-z^2) \epsilon}{Y^2 (2\epsilon+1)}\right) L_n(\hat{y},y)- \left(\frac{\epsilon}{Y^2}\right) \min\limits_{\beta\in[0,1]} \left\{ \frac{L_n(\hat{y}_{\beta},y)}{n}\right\}  \nn\\
&\leq O(1), \nonumber \\
&\frac{L_n(\hat{y},y)}{n}-\left( \frac{2 \epsilon+1}{1-z^2}\right) \min\limits_{\beta\in[0,1]} \left\{ \frac{L_n(\hat{y}_{\beta},y)}{n}\right\} \nn\\ &\leq \frac{2 \epsilon+1}{n\epsilon(1-z^2)}O(1) \leq O\left( \frac{1}{n\epsilon} \right),
\end{align}\normalsize
which is the desired bound.

Note that using $b = \frac{\epsilon}{Y^2}$, $a = \frac{(1-z^2) \epsilon}{Y^2
  (2\epsilon+1)}$ and $s = \left(\frac{Y^2}{2} + \frac{1}{4b}\right)$, we get
\begin{align*}
\mu = \frac{2+2\sqrt{1-4as}}{s} = \frac{4 \epsilon}{2\epsilon+1}\frac{2+2z}{Y^2}, \label{eq:mue}
\end{align*}\normalsize
after some algebra, as in the statement of the theorem.  This
concludes the proof of the theorem. $\Box$ \\

In the following lemma, we show that the order of the upper bound
using the KL divergence as the distance measure under the same
methodology cannot be improved by presenting an example in which the
bound on $b$ is of the same order as that given in the theorem.\\

\noindent
{\bf Lemma:} For positive real constants $a$, $b$ and $\mu$ which satisfies \eqref{desired_bound} for all $|y(t)| \leq Y$, $|\hat{y}_1(t)|\leq Y$ and $|\hat{y}_2(t)| \leq Y$ and $\lambda(t) \in [\lp, (1-\lp)]$, we require
\[
b \geq \frac{a}{4} + \frac{1}{16 \lp (1 -\lp)}.
\]\normalsize\\
\noindent
{\bf Proof:} Since the inequality in \eqref{desired_bound} should be
satisfied for all possible $y(t)$, $\hat{y}_1(t)$, $\hat{y}_2(t)$, $\beta$ and
$\lambda(t)$, the proper values of $a$, $b$ and $\mu$ should satisfy
\eqref{desired_bound} for any particular selection of $y(t)$,
$\hat{y}_1(t)$, $\hat{y}_2(t)$, $\beta$ and $\lambda(t)$. First we consider
$y(t)=\hat{y}_1(t)=Y$, $\hat{y}_2(t)=0$, $\beta=1$ and $\lambda(t)=\lp$ (or,
similarly, $y(t)=\hat{y}_1(t)=Y$, $\hat{y}_2(t)=-Y$ and
$\lambda(t)=\lp$). In this case, we have
\begin{align}
&a(Y - \lp Y)^2 \nn\\
&\leq - \ln (\lp + (1- \lp) e^{\mu (Y - \lp Y) \lp (1 - \lp) (-Y)}) \nonumber \\
& \leq - \lp \ln 1 - \mu (1- \lp)^2 \lp Y (1-\lp) (-Y) \label{tbound-1} \\
& = \mu (1- \lp)^3 \lp Y^2, \label{tbound-2}
\end{align}\normalsize
where \eqref{tbound-1} follows from the Jensen's Inequality for concave function $\ln(\cdot)$. By  \eqref{tbound-2}, we have
\begin{align}
\mu \geq \frac{a}{\lp(1- \lp)}. \label{main_bound-1}
\end{align}\normalsize

For another particular case where $\hat{y}_1(t)=Y$, $y(t)=\hat{y}_2(t)=0$, $\beta=1$ and $\lambda(t)=1/2$, we have
\begin{align}
a(- \frac{Y}{2})^2 - b (-Y)^2 & \leq - \ln (\frac{1}{2} + \frac{1}{2} e^{\mu (- \frac{Y}{2}) \frac{1}{4} (-Y)}) \nonumber \\
& \leq - \frac{1}{2}\mu \frac{Y^2}{8}, \label{tbound-3}
\end{align}\normalsize
where \eqref{tbound-3} also follows from the Jensen's Inequality. By \eqref{tbound-3}, we have
\begin{align}
b & \geq \frac{a}{4} + \frac{\mu}{16} \nonumber  \\
 & \geq \frac{a}{4} + \frac{a}{16 \lp (1 -\lp)}, \label{main_bound-2}
\end{align}
\normalsize
where \eqref{main_bound-2} follows from \eqref{main_bound-1}, which finalizes the proof. $\Box$

\section{Conclusion}
\label{sec:conclusion}
In this paper, we introduce a new and deterministic analysis of the
convexly constrained adaptive mixture of \cite{convex} without any
statistical assumptions on the underlying signals or any
approximations in the derivations. We relate the time-accumulated
squared estimation error of this adaptive mixture at any time to the
time-accumulated squared estimation error of the optimal convex
combination of the constituent filters that can only be chosen in
hindsight. We refrain from making statistical assumptions on the
underlying signals and our results are guaranteed to hold in an
individual sequence manner. We also demonstrate that the proof
methodology cannot be changed directly to obtain a better bound, in
the convergence rate, on the performance by providing counter
examples. To this end, we provide both the transient and steady state
analysis of this adaptive mixture in a deterministic sense without any
assumptions on the underlying signals or without any approximations in
the derivations.  \bibliographystyle{IEEEbib}
\bibliography{msaf_references}

\end{document}